\documentclass[journal=jacsat,manuscript=article]{achemso}

\usepackage[version=3]{mhchem}
\usepackage[numbers]{natbib}

\author{Nicolai T. Urban}
\affiliation{Max Planck Institute for Biophysical Chemistry, G\"ottingen, Germany}

\author{Matthew R. Foreman}
\affiliation{Blackett Laboratory, Department of Physics, Imperial College London, London SW7 2AZ}

\author{Stefan W. Hell}
\affiliation{Max Planck Institute for Biophysical Chemistry, G\"ottingen, Germany}

\author{Yonatan Sivan}
\email{sivanyon@bgu.ac.il}
\affiliation{Unit of Electro-optics Engineering, Faculty of Engineering Sciences, Ben-Gurion University of the Negev, P.O. Box 653, Israel, 8410501}

\title{Nanoparticle-assisted STED nanoscopy with gold nanospheres}

\begin{document}

\begin{abstract}
  We demonstrate stimulated emission depletion (STED) microscopy with 20 nm gold nanospheres coated by fluorescent silica. Compared with previous demonstrations of STED with a hybrid plasmonic fluorescent label, the current implementation offers a substantially smaller label and a better resolution improvement of up to 2.5-fold beyond the diffraction limit of confocal microscopy. This is achieved at approximately 2 times lower intensity than conventional STED based on dyes alone, and in an aqueous environment, demonstrating the relevance to bio-imaging. Finally, we also show, for the first time in this context, a 3-fold reduction in the rate of photobleaching compared to standard dye-based STED, thus, enabling brighter images.
\end{abstract}

\maketitle

\section{Introduction}
In recent years, there has been a rapid growth in the use of nanoparticles (NPs) for various applications. Metal nanoparticles~\cite{Giannini_chemrev} in particular, are finding applications in light concentration and harvesting in solar cell applications~\cite{plasmonic_photovoltaics}, sanitation~\cite{Halas-bubble1,Halas-bubble2}, sensing~\cite{Anker2008,Baaske2016}, surface-enhanced Raman scattering~\cite{SERS-book}, triggering chemical reactions~\cite{Halas_dissociation_H2,plasmonic-chemistry-Baffou}, super-resolution microscopy~\cite{plasmonic-SAX-ACS_phot,plasmonic-SAX-PRL,plasmonic-SAX-OE,plasmonic-SAX-rods-Ag,plasmonic_SIM_LPR}, along with many other applications. In the context of biological applications, they are being used for drug delivery~\cite{Xia_cages} and cancer therapy~\cite{Halas-PT_treat,Shaked-PT_treat}, for bio-diagnostics~\cite{Mirkin_Biodiagnostics}, as a nanoscale source of heat and acoustic waves~\cite{PT_imaging_Zharov,PT_PA_imaging_Zharov,PT_PA_imaging_Danielli}, as contrast agents in phase interferometry~\cite{Shaked-PT_imaging} and more.

Frequently, a metal NP or nanostructure is combined with a standard fluorescent label used in bio-imaging so as to create a  hybrid fluorescent label, with the goal of exploiting the strong fields occurring at the localized plasmon resonance (LPR) to boost the performance of the label. Specifically, such labels have been used for controlling the fluorescence lifetime and yield (also known as fluorescence engineering~\cite{Fluorescence_book}) and as a means to mitigate photobleaching~\cite{Stephane-sinks,Cang-NL-2013,Stefani_photobleaching,Galloway2014} and increase overall signal brightness~\cite{Renger_photobleaching}.

Recently, Sivan {\em et al.\ }proposed and demonstrated how such hybrid fluorescent labels can be used within the context of super-resolution microscopy. Specifically, within the context of stimulated emission depletion (STED) nanoscopy~\cite{Hell-review,Hell-Nobel}, it was shown theoretically~\cite{NP-STED-ACS-NANO,NP-STED-APL,NP-STED-book} that the near-field enhancement occurring at the LPR can be exploited to reduce the required intensity of the STED beam and to reduce the bleaching rates through the shortening of the lifetime of the excited singlet level; this technique is referred to as nanoparticle-assisted STED nanoscopy (NP-STED). The lowering of the STED intensity was demonstrated experimentally using 150~nm gold shell NPs with a STED beam at $\lambda_{\text{STED}} \approx 780$~nm~\cite{NP-STED-Experiment1}. However, the resolution level demonstrated in that work was limited to only a slight improvement with respect to the diffraction limit. In addition, a reduction of the bleaching rate had not been demonstrated in this context.

Clearly, realistic applications of NP-STED in bio-imaging will ensue only once substantial resolution improvement is demonstrated with a NP-STED label. In particular, more than a 2-fold improvement with respect to the diffraction limit is necessary in order to out-perform alternative approaches such as Structured Illumination Microscopy \cite{SIM}. This requires NP-STED labels which are small enough to be able to benefit from the enhanced spatial resolution, enjoy the maximum field enhancement levels, and minimize interference with the biological system. NP-STED would also benefit from extending the range of operating wavelengths and functions of the hybrid fluorescent labels.

In this article, we provide a step toward fulfilling these goals. Specifically, we demonstrate NP-STED with substantially smaller NPs and a shorter wavelength. A 2.5-fold improvement of resolution with respect to the diffraction limit of confocal microscopy is achieved, reaching resolutions of $\sim 100$~nm, while requiring $\sim 2$ times lower input intensity than needed for the standard dye. We also demonstrate, for the first time in this context, a reduced photobleaching rate induced by the presence of the metal NPs. Finally, we compare the performance to those reported previously~\cite{NP-STED-Experiment1} as well as more recently~\cite{NP-STED-Experiment2-UK}, and discuss the route towards realizing the full potential of NP-STED~\cite{NP-STED-ACS-NANO}. We also argue that even the existing implementation can be immediately combined with a wide range of imaging and treatment procedures utilizing metal NPs in a biological context.

\section{Principles of NP-STED}
\begin{figure}[htbp]
  \centering{\includegraphics[width=\columnwidth]{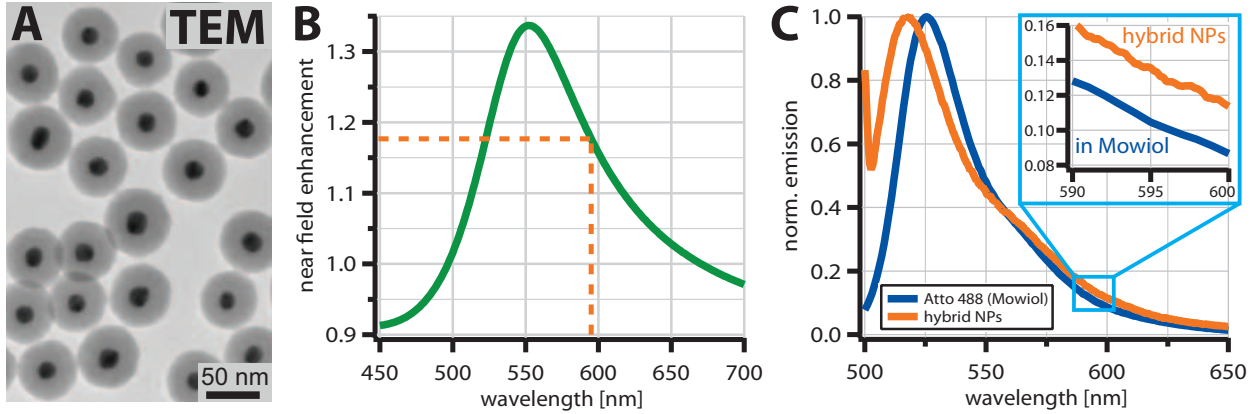}}
	%\centering{\includegraphics[scale=1.1]{STED_sphere_nanocomposix_12_14.eps}}
  \caption[]{Hybrid nanoparticles. (A) TEM image of the synthesized NPs, with 20~nm Au cores surrounded by 20~nm thick, dye-doped glass shells. (B) Spatially and orientationally averaged near-field intensity enhancement for the hybrid NPs immersed in water; highlighted is the enhancement value of 1.18 at the used STED wavelength ($595$~nm). (C) Normalized emission cross-section for both hybrid NPs and free Atto~488 in Mowiol; note the difference at ($595$~nm) of 0.14 (NPs) and 0.10 (Mowiol).\label{fig:NP}}
\end{figure}

The hybrid plasmonic-fluorescent labels used in this study consisted of 20~nm diameter gold spheres coated with a 20~nm glass shell doped with a standard green STED dye (Atto 488, which absorbs at $\sim 500$~nm, emits at $\sim 525$~nm and has a lifetime in the nanosecond range).  The particles, shown in Fig.~\ref{fig:NP}A, synthesized by Nanocomposix Inc., are substantially smaller than the gold shells used before~\cite{NP-STED-Experiment1}, as well as substantially easier to fabricate as compared with the thinner gold nanoshells required for optimal NP-STED performance~\cite{NP-STED-ACS-NANO,NP-STED-APL}. The STED intensity reduction due to the average near-field enhancement levels associated with this gold-core design, however, is substantially lower than those predicted for thin metal shells. The reason for this is that, unlike the rather uniform field experienced by dye {\em inside} a metal shell, the dye {\em around} a metal sphere experiences a field that decays as a function of the distance from the metal shell~\cite{Bohren-Huffman-book}.

%-->>>>>>>>>>>>>>>>>>>>>>>> Figure 1, TEM pics and enhancement <<<<<<<<<<<<<<<<<<<<<<<<<<<<<<<--%

Indeed, Fig.~\ref{fig:NP}B shows the near-field enhancement (defined as $\Gamma_I$ in~\cite{NP-STED-APL} and taken relative to a homogeneous Mowiol environment - see below) that was calculated using the parameters indicated above and averaged over the volume of the dye-doped glass coating for NPs in an aqueous environment. (In general, we denote the enhancement ratio of a quantity, NPs over standard dye, by $\Gamma$.)
It can be seen that at the local plasmon resonance~\cite{Giannini_chemrev}, $\lambda_{LPR} \approx 553$~nm, the (spatially) averaged near-field enhancement~\cite{Foreman_Sivan_PRB_2012} is about $\sim 1.34$. It was shown previously~\cite{NP-STED-APL,NP-STED-book} that such a simplistic calculation gives a reasonable estimate for the actual effective STED intensity reduction as calculated accurately with a doughnut shaped illumination (as \textit{e.g.}, performed in the initial studies~\cite{NP-STED-ACS-NANO}). In our setup, we opted for a STED wavelength of $\lambda_{\text{STED}} = 595$~nm, which is standard for Atto~488 and similar green dyes, but is slightly detuned from the plasmon resonance. At this wavelength the average near-field enhancement is $\sim 1.18$.

Besides overlapping with the STED wavelength, the plasmon resonance overlaps partially with the emission line of the dye, giving rise to a shortening of the fluorescence lifetime~\cite{Giannini_chemrev,Fluorescence_book}. In general, the decay rate enhancement depends strongly on position and the relative orientation of the emitter with respect to the metal surface. The cumulative signal emanating from a collection of such emitters typically exhibits a multi-exponential decay pattern~\cite{Giannini_chemrev,Smith_averaged_lifetime_calculation}. It has been shown, however, that it is frequently reasonable to approximate the decay curve with a few~\cite{Ruben_NJP} or even mono-exponential decay curves. In the latter case, one can define the decay rate enhancement, $\Gamma_k$, with respect to the free dye decay rate \cite{NP-STED-APL}. As long as the STED pulse duration is substantially shorter than the dye lifetime, there should be little influence on the STED efficiency of the nanoscope. If, however, the STED pulse duration approaches the dye lifetime or even becomes longer, then the STED efficiency will notably decrease. The effects of this can be (partially) compensated by using time-gated detection~\cite{STED_time_gating_theo,STED_time_gating_exp,NP-STED-APL}, with the caveat that time-gating will result in somewhat lower signal brightness. Under such time-gated collection, we have shown theoretically that the averaged near-field enhancement $\Gamma_I$ provides a good estimate of the STED intensity reduction that can be expected from the presence of the metal NP~\cite{NP-STED-APL,Foreman_Sivan_PRB_2012}.

Additionally, the fluorescence signal will further decrease due to the reduction of the apparent quantum yield (quenching) due to energy transfer to the metal, an effect enabled by the proximity of the fluorescent dye to the metal. For the NPs used in this study, the apparent average quantum yield was calculated to be about 0.15, \textit{i.e.\ }about 5 times lower than the quantum yield (0.8) of the free dye.

The lifetime shortening due to the NP has a positive effect as well, namely an associated reduction of bleaching. Indeed, the shortened time the molecule spends in an excited state reduces the overall number of chemical reactions that may lead to photobleaching (such as absorption to an excited state, or intersystem crossing to a triplet state~\cite{photo-bleaching1,photo-bleaching2,photo-bleaching3}). This effect was described by Enderlein~\cite{Enderlein_APL_2002,Enderlein_PCCP_2002}, and recently was demonstrated experimentally~\cite{Cang-NL-2013,Stefani_photobleaching} on single fluorescent molecules, showing a 5-fold reduction of the spatially-averaged bleaching rate. This was found to be in good agreement with the theoretical prediction (for monochromatic and uniform excitation) of the total photon yield being equal to the ratio of the emission and bleaching rates. A fuller analysis accounting for the spatial variation of both the lifetime shortening and near field enhancements~\cite{Galloway2014}, has further shown theoretically and confirmed experimentally that the resulting aggregate bleaching behavior can exhibit a multi-exponential behavior. An additional advantage associated with the decay rate enhancement, is that it allows higher excitation powers to be used, since the saturation threshold increases~\cite{Renger_photobleaching}. All these effects naturally enable much brighter signals, thus, compensating for the decrease of the quantum yield.

\section{Results}
\begin{figure}[htbp]
%   \floatbox[{\capbeside\thisfloatsetup{capbesideposition={right,top},capbesidewidth=0.5\textwidth}}]{figure}[\FBwidth]
	{\caption[]{Fluorescence decay of Atto 488 under different circumstances. (A) Confocal measurements show that the lifetime of Atto 488 is shortened considerably ($\tau=0.9$~ns, orange) when contained within the hybrid nanospheres, as opposed to within an immunohistochemically labelled cell ($\tau=2.8$~ns, dark blue) or suspended in water ($\tau=4.2$~ns, light blue). The shortened lifetime within the NPs matches well with the theoretical model ($\tau=0.95$~ns, black dashed). (B) When illuminated only with the STED beam (dark red), the hybrid NPs emit fluorescence with similar decay characteristics to regular excitation (orange dashed). If the STED power is high enough (yellow), the fluorescence is surpassed by metal luminescence with a very short lifetime, on the range of the system response function.}\label{fig:lifetimes}}
	{\includegraphics[width=0.5\columnwidth]{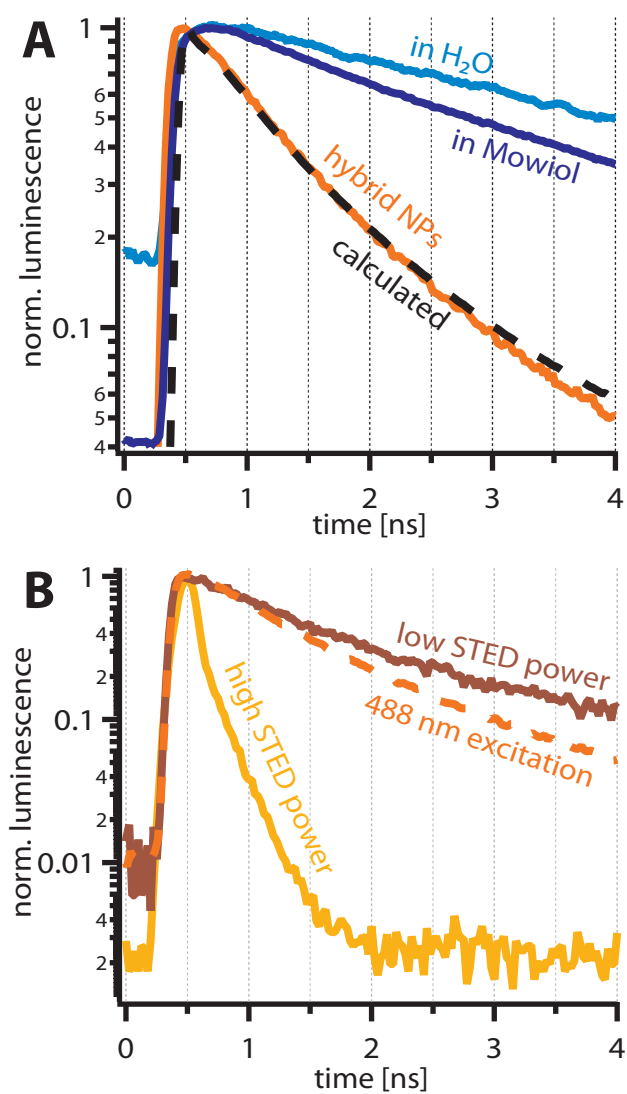}}
\end{figure}

%------------- fluorescence lifetime part -------------%
At first, we compared the fluorescence decay of the Atto 488 dye within the nanoparticle samples with measurements of free Atto 488 in water, as well as of Atto 488 used for immunohistochemically labeling fixated cells embedded in Mowiol. As can be seen in Fig.~\ref{fig:lifetimes}A, under 488 nm excitation, the measured lifetime of the dye contained within the nanoparticles was considerably shorter ($\tau=0.9$~ns) than when contained in fixated cells ($\tau=2.8$~ns) or when freely suspended in water ($\tau=4.2$~ns). The measured fluorescence decay within the NPs closely matched theoretical calculations of the lifetime shortening  ($\tau=0.95$~ns).

%-->>>>>>>>>>>>>>>>>>>>>>>> Figure 2, Lifetimes (confocal and STED-induced) <<<<<<<<<<<<<<<<<<<<<<<<<<<<<<<--%

\begin{figure}[htbp]
%   \floatbox[{\capbeside\thisfloatsetup{capbesideposition={right,top},capbesidewidth=0.5\textwidth}}]{figure}[\FBwidth]
	{\caption[]{Metal luminescence from the NPs under intense STED illumination. (A) When illuminating the NPs with only 595~nm light $<10$~mW, a weak fluorescence signal is visible (orange), which scales quadratically with the applied STED power. If the illumination intensity is increased above $\sim 10$~mW, then a second source of luminescence appears (yellow), which increases linearly in power and quickly drowns out the fluorescence. This metal luminescence displays a very short lifetime and a doughnut-shaped intensity profile (B). When recording proper STED images, the metal luminescence can be mostly discarded if the detection is time-gated (C), as opposed to using the entire detected signal (D). }\label{fig:luminescence}}
	{\includegraphics[width=0.5\columnwidth]{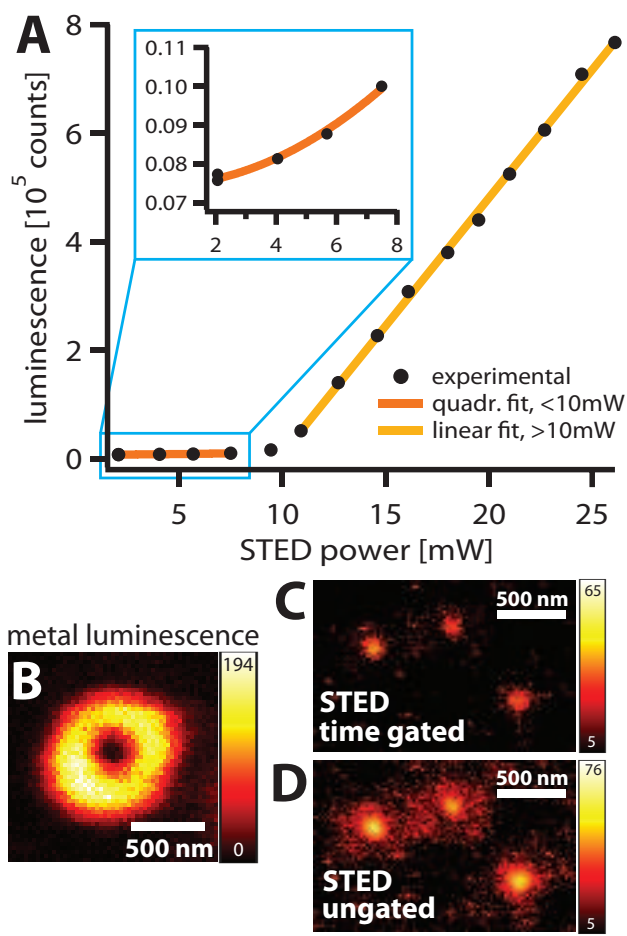}}
\end{figure}

%------------- Fast metal luminescence part -------------%
Besides the fluorescence, an additional source of luminescence became visible when illuminating the NPs with increasing STED power (Fig~\ref{fig:lifetimes}B). For simplicity, all stated laser powers correspond to the measured power in the back focal aperture of the objective lens. Using only the 595~nm STED beam as illumination, at low STED power ($<10$~mW) only a dim fluorescence signal could be detected originating from the NPs; it had a clear quadratic dependence on the incident intensity, hinting at a two-photon process. As the STED power was increased beyond $10$~mW, the dim fluorescence was quickly overwhelmed by a growing, fast luminescence signal, the lifetime of which was measured to be virtually indistinguishable from the instrument response function, \textit{i.e.\ }faster than $\sim 200$~ps (Fig.~\ref{fig:luminescence}A). The fast luminescence emanating from the metal NPs was clearly doughnut-shaped (Fig~\ref{fig:luminescence}B), as was observed in previous studies~\cite{NP-STED-Experiment1}. Furthermore, this luminescence signal was orders of magnitude brighter than the reflected signal from a plane mirror sample at identical illumination powers (data not shown) and could be observed over a wide range of wavelengths (from blue to green) at similar count rates. The power dependence of the fast luminescence at high illumination powers was linear (Fig.~\ref{fig:luminescence}A), suggesting a single-photon process. Due to the above, we can safely ascribe the unwanted luminescence at low powers to 2-photon fluorophore excitation by the STED beam, a known feature of the Atto dye family, while the luminescence at higher powers clearly originates from the metal. Indeed, metal luminescence is known to be very fast and to have a rather wide bandwidth~\cite{Mooradian_metal_luminescence,metal_shell_luminescence}. We do note, however, the difference with respect to the previous observation of metal luminescence in the context of NP-STED: previously,\cite{NP-STED-Experiment1} using a STED wavelength of $780$~nm, the metal luminescence seemed to scale with the square of the intensity, indicating two-photon absorption within the NP. For the much smaller NPs explored here, using a somewhat shorter wavelength of $595$~nm, the luminescence seems to originate from single-photon absorption.

Due to the very short duration of the metal luminescence signal, it was readily eliminated using time-gated detection (Fig.~\ref{fig:luminescence}C). By discarding the first 450~ps of the recorded signal, the signal originating solely from the fluorophores could be analyzed without being obstructed by the luminescence from the gold cores (Fig.~\ref{fig:luminescence}D). As previously noted, time-gating also served to eliminate a possible loss of STED efficiency, as the lifetime within the NPs was shortened in regards to the duration of the STED pulse. At 595~nm illumination powers higher than $\sim 20$~mW in the back focal aperture, the fast luminescence signal grew too strong to be ignored, as it began saturating the detector.

%-->>>>>>>>>>>>>>>>>>>>>>>> Figure 3, Metal luminescence & gating <<<<<<<<<<<<<<<<<<<<<<<<<<<<<<<--%

\begin{figure}[htbp]
  %\centering{\includegraphics[scale=0.8]{FWHM_figure.eps}} % P:/Writing_paper_latex/2016_Yonatan/
  %\centering{\includegraphics[scale=0.8]{P:/Writing_paper_latex/2016_Yonatan/fig4_FWHMs.eps}}
% 	\floatbox[{\capbeside\thisfloatsetup{capbesideposition={right,top},capbesidewidth=0.5\textwidth}}]{figure}[\FBwidth]
	{\caption[]{STED resolution enhancement for NP and control samples. Hybrid NPs recorded in (A) confocal and (B) STED mode (P$_{\text{STED}}^{(\text{inc})}=10$~mW). At identical STED powers, the hybrid NPs exhibited smaller FWHMs than the control samples. The data was fitted with an inverse square-root law% $d(P_{\text{STED}}^{(\text{inc})})=d_{\text{conf}}/{[ 1+P_{\text{STED}}^{(\text{inc})}/P_{\text{sat}}^*]^{1/2}}$
; the resulting saturation powers $P_{\text{sat}}^*$ are denoted in the figure. The ratio of saturation powers remained the same when measured using longer (200~ps, C) or shorter (50~ps, D) pulses.  \label{fig:FWHM}}}
	{\includegraphics[width=0.8\columnwidth]{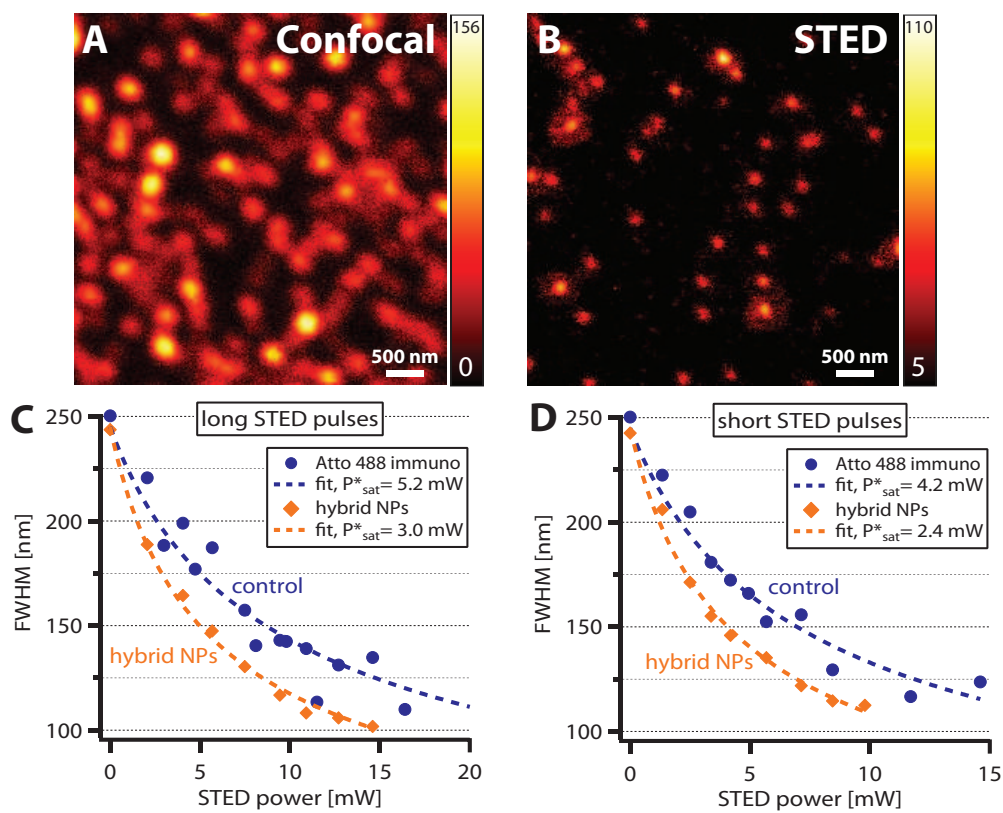}}
\end{figure}

%------------- FWHM/resolution improvement part -------------%
Next, we determined the achievable resolution of the hybrid NPs compared to the fixed-cell control samples for a given STED power P$_{\text{STED}}$, \textit{i.e.\ }we determined the different saturation intensities I$_{\text{sat}}$ of the standard and the hybrid NP dye. We estimated the achievable resolution by measuring the average FWHM (full-width at half-maximum) of the imaged structures (beads or filaments) at different STED powers, using time-gated detection. For simplicity, we will again use the measured incident STED power $P_{\text{STED}}^{(\text{inc})}$ in the back focal aperture of the objective lens as reference value. The resolution increase over confocal imaging due to STED (see Fig.~\ref{fig:FWHM}A,B) should follow an inverse square-root law, $d_{\text{STED}}\left(P_{\text{STED}}^{(\text{inc})}\right) \sim d_{\text{conf}}/\sqrt{1 + P_{\text{STED}}^{(\text{inc})}/P_{\text{sat}}^*}$, where the effective saturation level is given by $P_{\text{sat}}^* \equiv P_{\text{sat}} / \Gamma_I$ and $P_{\text{sat}} \sim h c / [\lambda_{\text{STED}} \sigma_{\text{em}}(\lambda_{\text{STED}}) T_{\text{STED}}]$,\footnote{Note that this expression differs by a factor of $(k_{\text{tot}} T_{\text{STED}})^{-1}$ from the expression for the saturation intensity used, \textit{e.g.}, in molecular physics, laser system etc., see discussion in~\cite{STED_OE_resolution_2010,NP-STED-book}. } with $\sigma_{\text{em}}$ the emission cross section, and $\lambda_{\text{STED}}$ and $T_{\text{STED}}$ the wavelength and duration of the STED pulse, respectively. Importantly, the expression incorporates the averaged near-field enhancement experienced by the dye due to the presence of the metal NP, but is independent of the total decay rate $k_{\text{tot}}$ and its enhancement factor $\Gamma_k$~\cite{NP-STED-APL}, as is discussed in previous studies~\cite{STED_OE_resolution_2010,NP-STED-book}.

In Fig.~\ref{fig:FWHM}C, the resolution with increasing STED power for both the NP and control experiments is shown, by plotting the measured FWHMs, averaged over several samples, as a function of the STED power P$_{\text{STED}}^{(\text{inc})}$. Compared to the control sample, the NPs required less STED power to achieve the same resolution enhancement. By fitting the data to a standard depletion curve, we determined that the saturation powers required for the NPs is reduced  by $\sim 1/1.75$ compared to the control. We repeated the measurements using shorter STED pulses ($\sim 50$~ps instead of $\sim 200$~ps), in case the shortened lifetime within the NPs negatively influenced the STED efficiency. Yet, the ratio of saturation powers was identical to the measurements using the long pulses (Fig.~\ref{fig:FWHM}D). We estimated the saturation intensities I$_{\text{sat}}$ of the samples to be about $8.1-10.0$~MW/cm$^2$ for the control sample and $4.6-5.8$~MW/cm$^2$ for the NPs. As compared to the theoretically predicted reduction in the required STED intensity of 1/1.18, the experimental result (1/1.75) is approximately 30\% better. This discrepancy is explained by the differences in the emission cross-section of the hybrid dye-NP label (see Fig.~\ref{fig:NP}C), however, other factors, such as NP size dispersion, roughness of the surface of the metallic core or the effect of the variation of the near-field enhancement $\Gamma_I$ as the doughnut-shaped STED beam is scanned across the NP~\cite{NP-STED-Experiment1} may be playing a role as well.

Notably, the results were fully reversible and repeatable; no photodamage of the scanned region of interest or the neighboring regions were observed, nor any signs of boiling the aqueous environment, such as bubbling, bleaching outside of the scanned areas, drifts, or movement of the NPs. Also, the variation of the observed resolution improvement between different NPs was much smaller compared with the previous experimental demonstration with Au shells~\cite{NP-STED-Experiment1} or the more recent example with Au rods~\cite{NP-STED-Experiment2-UK}. We were also able to utilize much higher STED intensities than before, possibly due to the lower metal volume hence yielding a higher threshold for damage from excess heating. Most importantly, unlike the initial demonstration of NP-STED where only a modest level of super-resolution was demonstrated~\cite{NP-STED-Experiment1,NP-STED-Experiment2-UK}, here (NP-)STED enabled a substantial ($>2$-fold) improvement of resolution with respect to the diffraction limit.

%-->>>>>>>>>>>>>>>>>>>>>>>> Figure 4, FWHMs (resolution enhancement, saturation intensities) <<<<<<<<<<<<<<<<<<<<<<<<<<<<<<<--%

%------------- Bleaching part -------------%
Finally, we compared the photostability of the hybrid NPs and the control sample for both confocal and STED measurements. For this, we continuously imaged small areas of the sample ($3~\mu\mbox{m} \times 3~\mu$m scan areas, pixel sizes of 100 nm (confocal) or 50 nm (STED), and pixel dwell-times between $\sim 10-100~\mu$s) until all or most of the fluorescence was bleached away, typically after $200-1000$ exposures. The chosen image parameters were identical for both the NP and the control samples, and after each imaging series we checked for possible sample drift (no drift was detected). As before, we time-gated the detection signal. The bleaching was quantified by plotting the remaining signal brightness (total photons over the entire scan area) over number of exposures and fitting the resulting signal decay with a bi-exponential function; the inverse decay constants of the bi-exponential fits were dubbed the `bleaching rates', with lower bleaching rates signifying more photostability. Such bi-exponential decay behavior has been observed before for various fluorescent dyes~\cite{photo-bleaching1,photo-bleaching3}. Both the fast and the slower bleaching rate showed near identical behavior for all the examined samples and imaging modalities. Thus, for simplicity, we will ignore the faster decay (with the higher variance) in favor of the slower one. As a measure of the improved photostability, we calculated the ratio of the inverse bleaching rates from the NP- and the control sample, denoted as $\Gamma_b$.

%-->>>>>>>>>>>>>>>>>>>>>>>> Figure 5, Bleaching <<<<<<<<<<<<<<<<<<<<<<<<<<<<<<<--%
\begin{figure}[htbp]
  %\centering{\includegraphics[scale=0.65]{bleaching_figure_v2.eps}} % P:/Writing_paper_latex/2016_Yonatan/
  \centering{\includegraphics[width=\columnwidth]{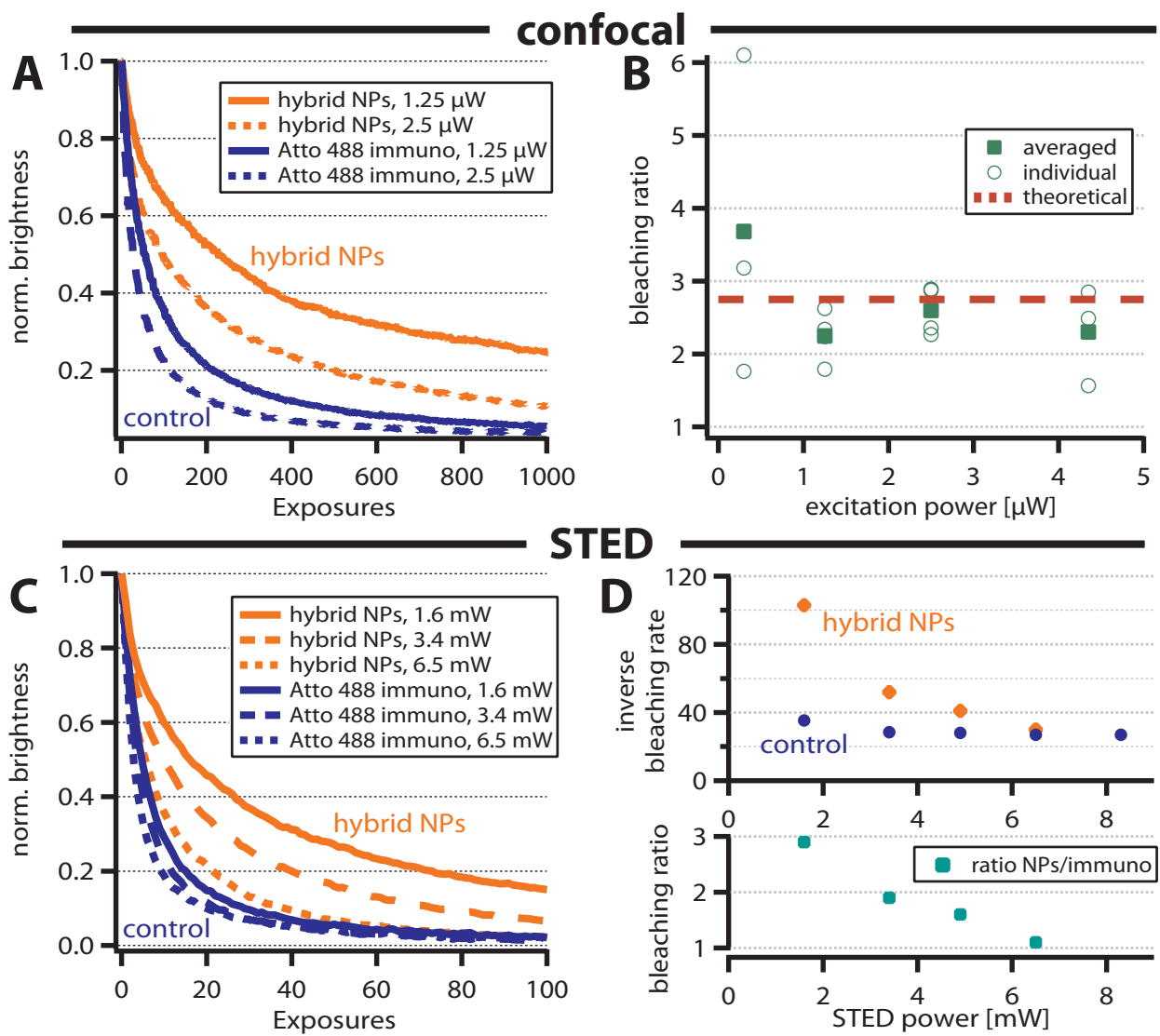}}
	{\caption[]{Bleaching behavior of hybrid NPs. (A) When illuminated with identical excitation powers (488~nm), the confocal signal recorded from the hybrid NPs (orange) bleaches away much more slowly than the control sample (blue). (B) The increased photostability of the NPs is mostly independent of the excitation power. (C) When imaging in STED mode (with both a 488~nm excitation and a 595~nm STED beam), the NPs (orange) again bleach more slowly than the control sample (blue). (D) The photostability is increased threefold at low STED powers, but as the STED intensity increases, the increased photostability of the NPs is counteracted by the thermal destruction of the dye.} \label{fig:bleaching}}
\end{figure}

Fig.~\ref{fig:bleaching} shows the fluorescence decay using both confocal and STED-enhanced imaging modalities. When imaging confocally, \textit{i.e.\ }without the STED beam, the control samples bleached away considerably faster than the dye protected within the hybrid NPs (Fig.~\ref{fig:bleaching}A). The bleaching ratio was found to be $\Gamma_b \sim 2-4$, seemingly irrespective of the utilized excitation power (Fig.~\ref{fig:bleaching}B), in good agreement with the theoretically predicted value of 2.75 (red dashed). We note that although the importance of the orientation of the dye molecule was known~\cite{Cang-NL-2013}, in many earlier treatments a fixed emitter orientation was considered, \textit{e.g.\ }due to preferential adsorption~\cite{Galloway2014}. In contrast, since we consider dye molecules doped in a silica coating, we allow for a random distribution of emitter orientations in turn requiring a further averaging step to be introduced in our modelling. When performing STED imaging, we again observed slower bleaching of the hybrid NPs (Fig.~\ref{fig:bleaching}C); at low STED intensities, the bleaching rate of the signal from the hybrid labels was about 3 times lower than the bare dye. As the STED power increased, however, the improved photostability of the hybrid labels decreased. Whereas the bleaching rate of the control sample remained roughly constant for increasing STED power, the bleaching of the hybrid NPs intensified, such that $\Gamma_b \to 1$ (Fig.~\ref{fig:bleaching}D). We ascribe the effect to damage occurring to the dye due to  heating of the metal and its environment~\cite{Galloway2014}, as this intense bleaching was also observed when illuminating the sample with only the STED beam, \textit{i.e.\ }in the absence of the excitation beam.

\section{Discussion}
The results detailed above provide a clear improvement with respect to the first proof-of-concept of NP-STED~\cite{NP-STED-Experiment1}. The current study demonstrates sub-diffraction resolution of $\sim 100$~nm with hybrid metal dielectric NP-STED labels, thus exceeding substantially the results presented in the initial study where only a modest super-resolution level was demonstrated. Crucially, this means resolution levels better than those attainable by other approaches, including Structured Illumination Microscopy. This is enabled by the 20-fold reduction of the label volume --- from $\sim 160$~nm gold shells~\cite{NP-STED-Experiment1} to particles which are $\sim 60$~nm diameter. In addition, our results show nearly a $\sim 2$-fold reduction of STED intensity required for achieving the same resolution with the current generation of organic labels.

Our study also confirms, for the first time in the context of STED, the prediction regarding the reduction of bleaching ensuing from the use of the hybrid metal-fluorescent labels~\cite{NP-STED-ACS-NANO} --- up to a 3-fold bleaching rate reduction was observed (in both confocal and STED modalities). This value is in good agreement with the theoretical prediction based on the theory presented previously~\cite{Galloway2014}, thus, showing that both additional bleaching reduction due to plasmonic-enhanced triplet lifetime shortening at the (resonant) STED wavelength~\cite{Stephane-sinks} as well as bleaching due to multi-photon processes is probably of lesser importance, at least for the dye used in this context.

Note that the bleaching reduction (without using STED) reported here is qualitatively similar to previous observations based on single molecules~\cite{Cang-NL-2013,Stefani_photobleaching}, however, in the confocal configuration it was obtained at substantially higher peak intensity --- about 2 orders of magnitude higher due to the spatial and temporal focused nature of the illumination. Moreover, in the presence of a STED beam, the associated intensities were an additional $3-4$ orders of magnitude higher. In the latter case, the depletion pulse, which enhances the decay from the excited level, modifies the quantitative description of the bleaching reduction~\cite{Galloway2014}, through addition of an extra decay pathway. Unfortunately, the bleaching reduction persists only for relatively low STED powers, as for higher powers, the heat generated in the metal seems to cause dye destruction~\cite{Galloway2014}.

Maybe the most important advance demonstrated in this study is that the imaging of the hybrid labels was performed in water (rather than in index matching oil, as was done so far~\cite{NP-STED-Experiment1,NP-STED-Experiment2-UK}). This highlights the possibility of applying NP-STED to the study of biological samples, specifically, to fixed cells. The applicability of NP-STED to live cell imaging, however, requires cell viability studies~\cite{PT_treatment,PT_damage,PT_damage-Sauer} to be carried out, in light of the transient temperature rise associated with the absorption in the metal.

Despite the important advances reported here, the performance improvement offered by the existing NP-STED hybrid labels is still limited. The major obstacle to achieving even better performance is NP miniaturization, and specifically, decreasing the metal volume. This would, on one hand, enable higher intensity reduction (and better maximal resolution). On the other hand, it would limit the temperature rise due to absorption in the metal, and thus, (besides the obvious reduced damage to the biological system), would extend the benefit of bleaching reduction to higher intensities. In that regard, the shorter STED wavelength we employed is advantageous: it allows use of gold spheres, or slightly elongated gold rods, as the LPR of these particles is sufficiently close to the STED wavelength used in our system. Whereas we used 20~nm Au cores, further miniaturization of the Au core and fluorescent shell is relatively easy to achieve. We note that the minimum size of a good NP-STED label can practically not be much below $10-20$~nm, since for fluorophore-metal separations of less than a few nanometers, the apparent quantum yield of the fluorophore will be too low due to fluorescence quenching. This could be compensated by the lower bleaching, however, and by the possibility to increase the excitation intensity~\cite{Renger_photobleaching}. Thus, we expect that particles which are 2--3 times smaller than those used in the current studies are realistic and should still yield comparable imaging performance.

A second concern is obtaining a more significant field enhancement, hence, STED intensity reduction. Further intensity reduction can be achieved by optimally matching the STED wavelength to the LPR band of the Au sphere~\cite{NP-STED-ACS-NANO,NP-STED-APL}. This can be achieved by either shortening the STED wavelength, or by using a STED dye with an emission line centered at even shorter wavelengths, such as AIE luminogen fluorophores~\cite{Sailing_STED}. Furthermore, the knowledge necessary for the selective positioning of the dye with respect to the metal surface is continuously being advanced, such that NP-STED labels with optimized performance, namely, higher field enhancement, lower quantum yield reduction, and moderate level of lifetime shortening, can be designed and synthesized quite easily. A step in this direction has already been demonstrated~\cite{NP-STED-Experiment2-UK}. Applying a similar procedure for the Au nanospheres used here, namely, replacing the thick dye-doped silica shell with flurophores linked at a specific distance from the metal surface can lead to up to 3-fold higher field enhancement. Alternatively, in the context of fixed cell imaging, so absent of living organisms, one could also employ Ag rods (considered toxic), for which the plasmonic performance exceeds that of Au by far, thus, enabling potential greater performance improvement in the NP-STED context.

Much higher intensity reduction (up to 50--100 fold) could be achieved with thin shells, such as those demonstrated previously~\cite{Fang_shells}; however, in that work, the cores were made of TiO$_2$, whereas for our purposes, a core material that can be doped with fluorophores is desired, \textit{e.g.}, silica. Similarly, one could also employ quantum dot cores, as STED has also recently been demonstrated with these emitters \cite{Qdot_STED}. The somewhat shorter wavelength used in the current study also poses less stringent constraints on the thickness of gold shells. Synthesis of the necessary thin shells is still challenging, but within reach.

Despite the limited improvement of performance, there are three avenues where the existing hybrid labels can find immediate use. One is to use the useful characteristics of the hybrid NPs to bolster existing methods, which do not currently use nanoparticles. The reduced saturation intensities could benefit parallelized STED schemes~\cite{parallelized-STED-Lounis,parallelized-STED}, where the lack of sufficient power can be a limiting factor in enabling the scan speed increase. Also, the reduced saturation intensities may also allow the application of STED to even thicker samples where the strong scattering reduces the incident STED intensities as one probes deeper into the tissue.

The second avenue entails using the hybrid NPs as multi-functional labels to enable the use of very different imaging methods. Specifically, on one hand, operating in STED mode, one can benefit from the high spatial resolution levels at the emission wavelength; in the absence of a STED beam, and for sufficiently intense excitation, one can benefit from the photothermal/photoacoustic response of the gold core~\cite{PT_imaging_Zharov,PT_PA_imaging_Zharov,PT_PA_imaging_Danielli} and the strong, spectrally-narrow signal achievable above the nanolasing (also known as spasing) threshold enabled by the presence of feedback from the plasmonic cavity (or cavities)~\cite{SPASER_Zharov}; in the same modality, one can perform either bleaching-free imaging at enhanced resolution at the excitation wavelength via LPR-based Structured Illumination~\cite{plasmonic_SIM_LPR} using low intensities, quantitative phase contrast microscopy via the photothermal effect~\cite{Shaked-PT_imaging} using moderate intensities, or plasmonics-based saturated excitation (SAX) microscopy~\cite{plasmonic-SAX-ACS_phot,plasmonic-SAX-PRL,plasmonic-SAX-OE,plasmonic-SAX-rods-Ag,SUSI} using intense monochromatic illumination. One can also correlate the super-resolved STED signal with 2-photon imaging~\cite{B_Yakar_NL_2007} using moderate to high intensities. Ultimately, the hybrid fluorescent labels studied here can serve towards correlated light electron imaging~\cite{CLEM}.

The third avenue, and potentially the most important one, relies on enhancing the many existing bio-related applications of metal NPs with a significant level of super-resolution imaging, creating a versatile bio-tool. These applications include, for example, selective treatment of cancer cells via hyperthermia~\cite{Halas-PT_treat,Shaked-PT_treat}, membrane perforation and gene therapy treatments~\cite{Meunier_cell_perforation}, stimulation and monitoring signalling in neurons~\cite{Meunier_neuron_signalling}, tweezing~\cite{Gu_NPs} and drug delivery~\cite{Xia_cages,Xia_analyst}. A widely unexplored and promising area, is biodiagnostics~\cite{Mirkin_Biodiagnostics} based on spherical nucleic acids~\cite{Mirkin_SNA}, which, again, are quite similar to the NPs used in this study. The wide variety of associated applications includes protein targeting, selective colorimetric detection of polynucleotides~\cite{Mirkin_Science_1997}, scanometric DNA array detection~\cite{Mirkin_DNA_array_detection} and intracellular gene regulation~\cite{Mirkin_gene_regulation}, to name just a few. These successful applications could all benefit greatly from the substantial improvement of the imaging resolution offered by NP-STED.

\section{Methods}
%------------- Sample preparation part -------------%
{\bf Sample preparation.} The hybrid nanoparticles were diluted in pure water, then sonicated to ensure that no agglomeration occurred, and finally applied to coverslips coated with a thin Poly-L Lysine layer (for increased stickiness of the surface). The coverslips were then briefly rinsed with pure water (to remove excess NPs and dye) and subsequently dried, before re-immersing them in a small drop of water and affixing the coverslip to the microscope slide. The particles had a density of approximately 5--10 beads per 1 $\mu\text{m}^2$, and only seldomly were there clusters of agglomerated beads. This way, we avoided tweezing effects and NP diffusion, which occurred previously. The control samples consisted of MeOH-fixated Vero cells, immunohistochemically stained with a primary anti-vimentin antibody (Sigma V6389, 1:100 dilution) and a secondary Atto-488 sheep/anti-mouse antibody ($1:50$ dilution) and mounted in Mowiol (with a refractive index of $n = 1.45$).

%------------- Setup part -------------%
{\bf STED nanoscope.}  All experiments were performed on a custom-built STED nanoscope~\cite{STED-595-Urban-set-up} optimized for use with green fluorophores. For excitation, a 488~nm pulsed diode laser (PicoTA, Toptica Photonics, Graefelfing, Germany) was used with pulse lengths of $\approx 110$~ps and typical laser powers between 1--10 $\mu$W in the back focal aperture of the objective lens. The STED beam was created using a Ti:sapphire laser (80~MHz; MaiTai, Spectra-Physics, Darmstadt, Germany), which was frequency-shifted using an optical parametric oscillator (APE, Berlin, Germany) to a wavelength of 595~nm. The initially femtosecond pulses were pre-stretched using SF6 glass rods before being stretched further by either a short (25m) or a long (110m) polarization maintaining glass fiber (OZ Optics, Ottawa, Canada), resulting in pulse widths of either approximately 50~ps or 200~ps, respectively. The typical power of the STED beam in the back focal aperture was between $1-25$~mW. For our system, which employs a 80~MHz pulse repetition rate, a rough estimate
shows that these power levels correspond to $\sim 17 - 425$~MW/cm$^2$.

The beams were overlapped using custom-made dichroic mirrors and focused using a 1.3~NA glycerol immersion objective lens (63x, PL APO, CORR CS; Leica, Wetzlar, Germany). The fast scan axis was performed using resonant mirror scanning (15 kHz; SC-30, EOPC, Glendale, NY) and the slow axis using a piezo stage (P-733; Physik-Instrumente, Karlsruhe, Germany). The fluorescence passed through a bandpass filter (535/60) and was recorded using a single-photon avalanche photo diode (SPAD, PDM series, Micro Photon Devices, Bolzano, Italy); images were recorded using the TTL-output of the SPAD, whereas the fluorescence decay measurements used the fast NIM-output with 35~ps time resolution and were counted using a time-correlated single-photon counting module (SPC150N, Becker \& Hickl, Berlin, Germany). A second detection window (450/60) could be used simultaneously for observing light outside of the Atto 488 fluorescence spectrum (\textit{i.e.\ }metal luminescence).

%------------- Measurement details part -------------%
{\bf Measurement details.} All STED images were recorded using time-gated detection, separating the first 450~ps after onset of the excitation pulse from the rest of the signal, in order to distinguish fluorescence from any occurring metal luminescence. Nevertheless, when imaging the hybrid nanoparticles, we limited the applied STED power to $<15$~mW, to avoid saturating the detector, risk imaging artifacts due to residual metal luminescence and/or damage to the NPs due to heating. This limited the achievable resolution to $>$100~nm (thus not sufficient to fully resolve the 60~nm hybrid nanoparticles), instead of the better than $\sim 50-70$~nm resolution, which the STED nanoscope is capable of (when imaging control samples).
The lifetime of the dye under the various circumstances was determined by fitting the fluorescence decay using a routine implemented in Matlab (Mathworks, MA, USA), which iteratively reconvolved the instrument response function (IRF) with a mono- or bi-exponential decay function and then optimized the parameters. The IRF was acquired by reflecting the excitation or STED-beam, respectively, off of a plane mirror sample and recorded using the same detection path as for the experiments. The IRF displayed a timing resolution of $\sim 110$~ps FWHM for the 488~nm excitation beam and $\sim 200$~ps FWHM for the 595~nm STED beam.

%------------- Theoretical calculations/modelling -------------%
{\bf Theoretical calculations and modelling} To determine the electric field distribution for the core-shell geometry NPs, standard Mie theory~\cite{Aden1951} was used assuming an incident circularly polarised plane wave. Nominal values, provided by Nanocomposix Inc.\ of 18.8~nm and 20.5~nm were used for the inner core diameter and shell thickness, respectively. In all calculations, the properties of gold were taken from~\cite{Palik-book} whilst the refractive index of the silica coating and the aqueous host were taken as $1.4584$ and $1.3324$. The near field enhancement $\Gamma_I$ was determined by taking the ratio of the resulting intensity distribution as compared to that of a plane wave in a homogeneous Mowiol environment ($n = 1.45$), so as to match experimental controls. Volume averages were taken within the NP coating only. Results were verified using COMSOL 4.3a.

Position dependant radiative Purcell factors, $F_r = P_r/P_0$ for dipole emitters oriented both radially and tangentially with respect to the core were calculated using a method based on accepted theory\cite{Moroz2005}. Specifically, the integrated power flow through a spherical surface in the far field ($P_r$), was compared to that of a free emitter embedded in Mowiol ($P_0$), for different emission wavelengths $\lambda_{\text{em}}$. The latter was found using the Larmour formula. Similarly, non-radiative Purcell factors ($F_{nr} = P_{nr}/P_0$) followed by calculating the power absorbed in the metal core, $P_{nr}$. The lifetime trace in Fig.~\ref{fig:lifetimes} was then calculated by considering the number of photons emitted by a dipole of given orientation $\mathbf{p}$ and position (neglecting photobleaching) which, as discussed previously~\cite{Smith_averaged_lifetime_calculation}, is proportional to $N(\mathbf{r},\hat{\mathbf{p}},t) \sim |\hat{\mathbf{p}}\cdot \mathbf{E}| \exp[-(F_r + F_{nr} + \eta_0^{-1} -1) t/\tau_0]$, where $\eta_0 = 0.8$ ($\tau_0$) is the quantum yield (lifetime) of a free emitter, caret notation denotes a unit vector and $\mathbf{E}$ is the electric field at the position of the dipole upon illumination of the NP with a plane wave. Assuming that the dye molecules were oriented randomly over the full $4\pi$ solid angle $\Omega$, the total collected signal was found by integrating over the volume of the shell and averaging over the orientation. Each spectral component was further weighted according to the Atto 488 emission spectrum $w(\lambda_{\text{em}})$ such that the fluorescence intensity scales as
\begin{align}\label{eq:avg}
I(t) \sim \frac{1}{4\pi} {\int_0^\infty \int_{\Omega} \int_V  w(\lambda_{\text{em}}) N_T(\mathbf{r},\hat{\mathbf{p}},t,\lambda_{\text{em}})  d\mathbf{r}d\hat{\mathbf{p}} d\lambda_{\text{em}}}.
\end{align}
Finally, the resulting fluorescence trace was convolved with the measurement IRF. We note there were no free parameters in our calculation. The effective lifetime was extracted using a mono-exponential fit.

Since the lifetime of differently oriented and positioned emitters within the NP is modified to varying extents, the relative competition between radiative decay, non-radiative decay and photobleaching also varies. The effective bleaching rate of a given emitter can be determined as in~\cite{Galloway2014}, however, the resultant bleaching curves (c.f. Fig.~\ref{fig:bleaching}) will, similarly to the lifetime traces, be formed of the superposition of multiple exponential curves. Using a normalized time coordinate, the decay constants in the exponents are given by $\alpha \Gamma_I / [F_r + F_{nr}+ \eta_0^{-1} - 1]$, where $\alpha = |\hat{\mathbf{p}}\cdot\hat{\mathbf{E}}|^2/|\hat{\mathbf{p}}\cdot\hat{\mathbf{E}}_0|^2$ describes the relative polarization rotation from the NP. The relative weighting of the individual bleaching curves depends on the local near field intensity enhancement $\Gamma_I$ and the effective quantum yield $\eta = \eta_0 F_r / (F_r + F_{nr})$ of the emitter~\cite{Galloway2014}. Volume, orientation and wavelength averages were taken similarly to Eq.~\eqref{eq:avg}. To parallel the experimental treatment, mono-exponential fits were used to determine effective decay rates and hence the theoretical bleaching ratio $\Gamma_b$.

% \bibliography{my_bib}
%\bibliography{D:/MyDocs/Research/my_bib}
% \bibliography{C:/Users/Yonatan/Documents/Research/my_bib}

\providecommand{\latin}[1]{#1}
\providecommand*\mcitethebibliography{\thebibliography}
\csname @ifundefined\endcsname{endmcitethebibliography}
  {\let\endmcitethebibliography\endthebibliography}{}

\end{document}